\documentclass[aps,twocolumn]{revtex4}

\usepackage{graphicx}
\usepackage{dcolumn}
\usepackage{bm}

\begin{document}

\title{Phase transitions in the one-dimensional ionic Hubbard model}

\author{Myung-Hoon Chung}
\email[]{mhchung@hongik.ac.kr}
\affiliation{College of Science and Technology, Hongik University,
Sejong 339-701, Korea}

\date{\today}

\begin{abstract}
We study quantum phase transitions by measuring the bond energy, the number density,
and the half-chain entanglement entropy in the one-dimensional ionic 
Hubbard model. By performing the infinite density matrix renormalization
group with matrix product operator, we obtain ground states 
as the canonical form of matrix product states. Depending on the chemical potential and
the staggered potential, the number density and the half-chain
entanglement entropy shows clear signatures of the Mott transition.
Our results confirm the success of the matrix product operator method for
investigation of itinerant fermion systems.
\end{abstract}


\maketitle

\section{Introduction}

Intrinsic characteristics of quantum
entanglements \cite{Osterloh, Osborne}
are essential elements in the study of quantum phase transitions \cite{Sachdev}.
In recent years, there has been a great deal of interest in entanglement entropy \cite{Eisert},
which is a quantum phase transition marker \cite{Cha} in the interacting
spin \cite{Tagliacozzo, Pollmann, Pirvu}, boson \cite{Pino}, 
and fermion \cite{Gu, Larsson, Iemini, Cha18} systems.

Some cold atoms can obey the same statistical rules as electrons if
atoms with an odd number of neutrons are fermions \cite{Parsons}. 
Thus, some cold atoms \cite{Cheuk} in an
optical lattice can mimic the behavior of electrons in a real solid material.
Since cold fermion systems are invented in quasi one-dimensional
optical lattices \cite{Schreiber}, the one-dimensional fermion
Hubbard model can be realized physically. 
The one-dimensional fermion Hubbard model \cite{Essler} was analytically solved
and became a prototype of a playground, where 
theoretical results could be compared to physical reality.

\begin{figure}
\includegraphics[width= 8.0 cm]{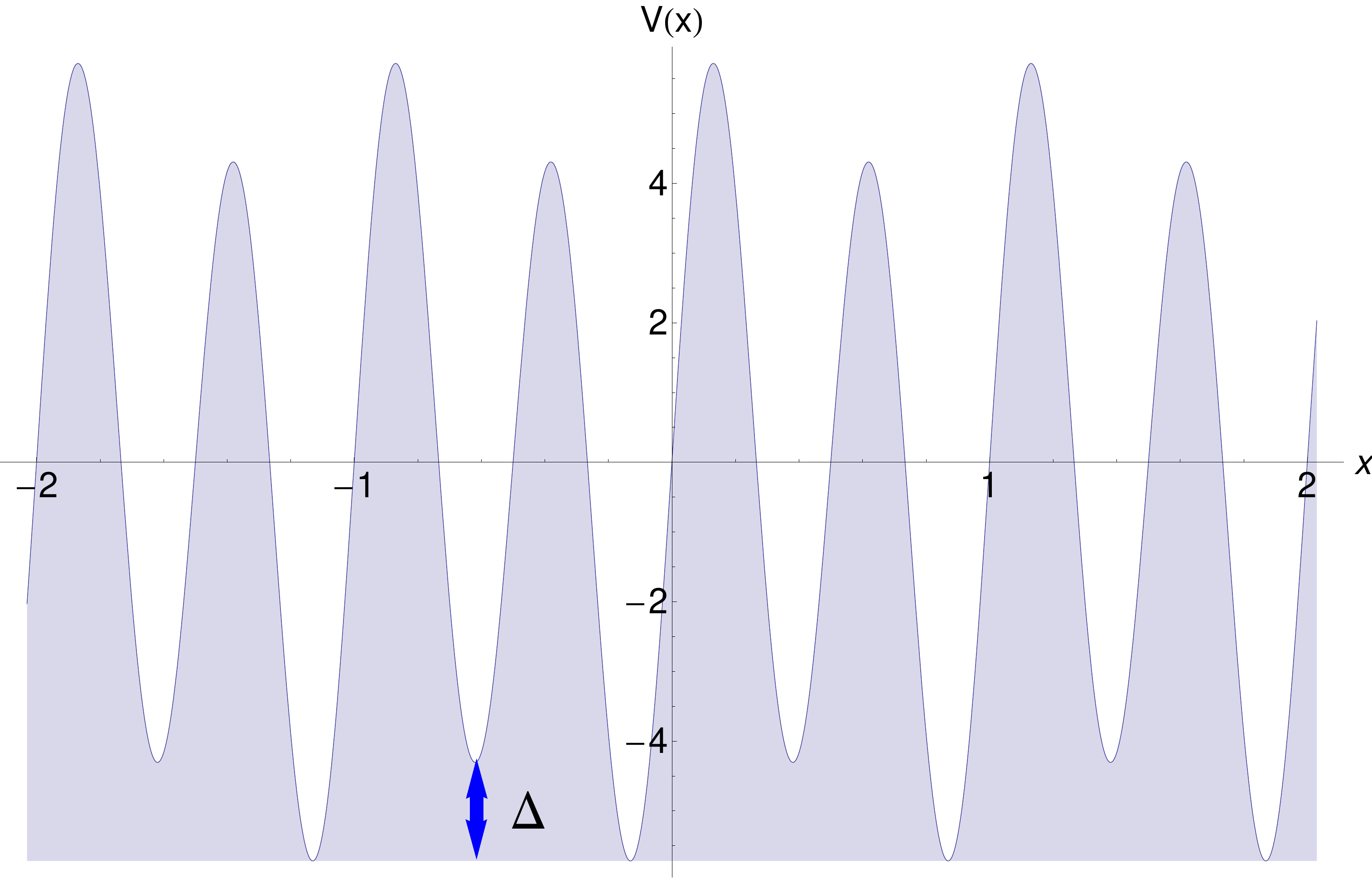}
\caption {The shape of the potential energy made by two wavelength lasers
$\lambda$ and $2\lambda$. Explicitly, we plot $V(x) = 5 \sin (4 \pi
x) + \sin( 2 \pi x)$. We notice the staggered potential $\Delta$.  \label{fig:fig1} }
\end{figure}

In this paper, motivated by cold atoms, we study the one-dimensional Hubbard model with
an energy offset between even and odd lattice sites. The main process in 
this study aims to see the role of different potential energies, which
is made by the second laser disturbance in the one-dimensional
cold atom systems. In fact, we impose the potential energy as
\begin{equation}
V(x) = V_{\lambda} \sin (kx) + V_{2\lambda} \sin(\frac{k}{2} x),
\label{eq:V}
\end{equation}
using the second laser of the doubly long wavelength. For example,
we present the staggered potential with typical values as shown in Fig.
\ref{fig:fig1}. 

The ionic Hubbard model \cite{Torbati} stemmed from this staggered potential.
There have been many investigations into the ionic Hubbard model at half filling,
by using 
dynamical mean field theory \cite{Bag, Lin},
determinant quantum Monte Carlo \cite{Bouadim},
exact diagonalization and finite density matrix renormalization group \cite{Kampf}.

The purpose of this paper is to find a quantum phase diagram
in the one-dimensional ionic Hubbard model away from half filling.
To do this, we propose the ansatz
of matrix product state (MPS) for the ground state, and perform 
the infinite density matrix renormalization group (iDMRG) \cite{McCulloch} method
with matrix product operator (MPO). 
By changing the chemical potential and
the staggered potential, we calculate the entanglement entropy, 
which shows quantum phase transition. The occupation number density also indicates the
transition consistently. We determine the phase diagram in the parameter space.

\section{Model and method}

Introducing the staggered one-particle potential $\Delta$ into the
Hubbard model, we consider a simple quantum system of interacting
fermions on a one-dimensional lattice. The Hamiltonian for this
model is written as
\begin{eqnarray}
H &=& -t\sum_{\langle i j
\rangle}(c^{\dagger}_{i\uparrow}c_{j\uparrow} +
c^{\dagger}_{j\uparrow}c_{i\uparrow} +
c^{\dagger}_{i\downarrow}c_{j\downarrow} +
c^{\dagger}_{j\downarrow}c_{i\downarrow}) \nonumber \\
& & +U\sum_{i}(n_{i\uparrow}-\frac{1}{2})(n_{i\downarrow}-\frac{1}{2})  
-\mu\sum_{i}(n_{i\uparrow}+n_{i\downarrow}) \nonumber \\
& & +\frac{\Delta}{2}\sum_{i \in \text{even}}(n_{i\uparrow}+n_{i\downarrow}) 
    -\frac{\Delta}{2}\sum_{i \in \text{odd}} (n_{i\uparrow}+n_{i\downarrow})  ,
\label{eq:H}
\end{eqnarray}
where $\langle i j \rangle$ represents the nearest-neighbor hopping in
a one-dimensional lattice, and $n_{i\uparrow}$ and
$n_{i\downarrow}$ are the spin-up and the spin-down number
operator, respectively. We let the hopping strength $t$ be 1 and
vary the strengths of the parameters: the on-site Coulomb
repulsion $U$, the chemical potential $\mu$, and the staggered potential 
$\Delta$. The role of the chemical
potential is to control the number of fermions in the system.
For the case of $\Delta=0$, the exact solution
of the original Hubbard model is known \cite{Yang}.

In order to handle $\mu$ and $\Delta$, we introduce two linear combinations 
$\mu_{e}$ and $\mu_{o}$ for even and odd sites, denoting 
\begin{equation}
\mu_{e} \equiv \mu - \frac{\Delta}{2}, ~~~ \mu_{o} \equiv \mu + \frac{\Delta}{2}.
\label{eq:mu}
\end{equation}
Because the Hamiltonian has the two-site translational
symmetry as shown in Fig. \ref{fig:fig1}, we apply iDMRG with
unit cell of two sites to determine the ground state. Our strategy is
to use MPO acting on MPS.

In relation to the basis, we set the physical index $\sigma_{i}$ in
the MPS. The state on the $i$-th site is
represented by $\sigma_{i}=(\alpha_{i},\beta_{i})$, such as
$0=(0,0)$ for the empty, $1=(0,1)$ for the spin-down, $2=(1,0)$
for the spin-up, and $3=(1,1)$ for both of the spin-up and the
spin-down. Thus, the physical index runs from $0$ to $3$. The
basis of the Fock space is written in terms of creation
operators $c^{\dagger}_{i\uparrow}$ and
$c^{\dagger}_{i\downarrow}$ as follows:
\begin{eqnarray}
| \sigma_{0} \cdots \sigma_{L-1} \rangle &=&
(c^{\dagger}_{0\uparrow})^{\alpha_{0}}
(c^{\dagger}_{0\downarrow})^{\beta_{0}}
(c^{\dagger}_{1\uparrow})^{\alpha_{1}}
(c^{\dagger}_{1\downarrow})^{\beta_{1}} \nonumber \\
& &\cdots
(c^{\dagger}_{L-1\uparrow})^{\alpha_{L-1}}
(c^{\dagger}_{L-1\downarrow})^{\beta_{L-1}}  |0 \rangle ,
\label{eq:basis}
\end{eqnarray}
where $\alpha_{i} (\beta_{i}) = 0$ or $1$ means that there is a
spin-up (down) fermion vacancy or occupancy at the $i$-th site,
respectively. It is important to maintain the order of fermions 
in the state of the Fock space to deal with the negative
sign caused by fermion exchange. We adopt the order of
spin-up first and then spin-down as described above.

In the process of iDMRG, we have to find the MPO for the
Hamiltonian of Eq. (\ref{eq:H}).
We convince that the MPO of the $i$-th site is
indeed given by the following form:
\begin{equation}
M_{i}= \left( \begin{array}{cccccc} 1 & c^{\dagger}_{i\uparrow} &
c^{\dagger}_{i\downarrow} & c_{i\uparrow} & c_{i\downarrow} & b_{i} \\
0 & 0 & 0 & 0 & 0 & -t c_{i\uparrow}   \\
0 & 0 & 0 & 0 & 0 & -t c_{i\downarrow} \\
0 & 0 & 0 & 0 & 0 &  t c^{\dagger}_{i\uparrow}    \\
0 & 0 & 0 & 0 & 0 &  t c^{\dagger}_{i\downarrow}  \\
0 & 0 & 0 & 0 & 0 & 1
\end{array} \right)
\end{equation}
where
\begin{equation}
b_{i} = U(n_{i\uparrow}-\frac{1}{2})(n_{i\downarrow}-\frac{1}{2})
-\bar{\mu} (n_{i\uparrow}+n_{i\downarrow}),
\end{equation}
\begin{equation}
\bar{\mu} = \left\{ \begin{array}{cc} \mu_{e} & \text{for }i=\text{even,} \\
                                      \mu_{o} & \text{for }i=\text{odd.}  \\
                    \end{array} \right.
\end{equation}
Here, we omit to present the boundary operators. We note that the
MPO for even sites is slightly different from the MPO for odd sites, where
$\mu_{e}$ must be replaced with $\mu_{o}$.
Because $c$ and $c^{\dagger}$ are fermion operators, the commutation relations between $M_{i}$
and $c^{\dagger}_{j\sigma}$ ($i \neq j,~\sigma=\uparrow,\downarrow$) are given by $M_{i}c_{j\sigma}^{\dagger} = c_{j\sigma}^{\dagger}M_{i}^{*}$, where
\begin{equation}
M_{i}^{*} = \left( \begin{array}{cccccc} 
1 & -c^{\dagger}_{i\uparrow} & -c^{\dagger}_{i\downarrow} 
& -c_{i\uparrow} & -c_{i\downarrow} & b_{i} \\
0 & 0 & 0 & 0 & 0 & t c_{i\uparrow}   \\
0 & 0 & 0 & 0 & 0 & t c_{i\downarrow} \\
0 & 0 & 0 & 0 & 0 &  -t c^{\dagger}_{i\uparrow}    \\
0 & 0 & 0 & 0 & 0 &  -t c^{\dagger}_{i\downarrow}  \\
0 & 0 & 0 & 0 & 0 & 1
\end{array} \right)
\end{equation}
When the Hamiltonian acts on the basis, the sign caused by fermion exchange 
must be taken care of such as
\begin{eqnarray}
M_{0}\cdots M_{L-1}
(c^{\dagger}_{0\uparrow})^{\alpha_{0}}
(c^{\dagger}_{0\downarrow})^{\beta_{0}}
(c^{\dagger}_{1\uparrow})^{\alpha_{1}}
(c^{\dagger}_{1\downarrow})^{\beta_{1}} \nonumber \\
\cdots (c^{\dagger}_{L-1\uparrow})^{\alpha_{L-1}}
(c^{\dagger}_{L-1\downarrow})^{\beta_{L-1}}  |0 \rangle ~~~~~~~\nonumber \\
~~~ = M_{0} (c^{\dagger}_{0\uparrow})^{\alpha_{0}}
        (c^{\dagger}_{0\downarrow})^{\beta_{0}}
~ \bar{M}_{1} (c^{\dagger}_{1\uparrow})^{\alpha_{1}}
        (c^{\dagger}_{1\downarrow})^{\beta_{1}} \cdots \nonumber \\
~ \bar{M}_{L-1} (c^{\dagger}_{L-1\uparrow})^{\alpha_{L-1}}
        (c^{\dagger}_{L-1\downarrow})^{\beta_{L-1}}   |0 \rangle , ~~~~~~~~
\label{eq:sign}
\end{eqnarray}
where the total number of fermions in the left side of the $i$-th site, 
$N_{i} \equiv \sum_{k=0}^{i-1}(\alpha_{k}+\beta_{k})$, determines the $6 \times 6$ matrix: 
\begin{equation}
\bar{M}_{i} = \left\{ \begin{array}{cc} M_{i}   & \text{for }N_{i}=\text{even,} \\
                                      M_{i}^{*} & \text{for }N_{i}=\text{odd.}  \\
                    \end{array} \right.
\end{equation}
In order to manage the sequence of Eq. (\ref{eq:sign}), 
we note that the even and odd structure of the fermion
numbers $(E~~O)$ is kept recursively by $e \equiv \{ |0\rangle , c^{\dagger}_{\uparrow}c^{\dagger}_{\downarrow} |0\rangle \}$ 
and $o \equiv \{ c^{\dagger}_{\uparrow}|0\rangle , c^{\dagger}_{\downarrow} |0\rangle \}$ 
such as 
\begin{eqnarray}
(E~~O) \left( \begin{array}{cc} e & o \\
                                o & e \\
                    \end{array} \right)
       \left( \begin{array}{cc} e & o \\
                                o & e \\
                    \end{array} \right) \cdots
       \left( \begin{array}{cc} e & o \\
                                o & e \\
                    \end{array} \right)  \nonumber \\
=(E~~O) \left( \begin{array}{cc} e & o \\
                                o & e \\
                    \end{array} \right) \cdots
       \left( \begin{array}{cc} e & o \\
                                o & e \\
                    \end{array} \right)  .
\label{eq:eo}
\end{eqnarray}
Using the relation of Eq. (\ref{eq:eo}), 
we double the MPO for the $i$-th site, $\bar{M}_{i}$, which
is modified into $12 \times 12$ matrix, $D_{i}$, such as 
\begin{equation}
       D_{i} e = 
       \left( \begin{array}{cc} M_{i}e & 0 \\
                                0 & M_{i}^{*}e \\
                    \end{array} \right) ,  ~~
       D_{i} o = 
       \left( \begin{array}{cc} 0 &  M_{i}o\\
                                M_{i}^{*}o  & 0\\
                    \end{array} \right) ,
\end{equation}
where $M_{i}^{*}$ is used in the second row because there are odd number of fermions
in the left side of the $i$-th site.
By including the $4$-dimensional physical index, we present the effective MPO
as the graphical diagram with the four-leg tensor $D$ having 
$4 \times 12 \times 12 \times 4$ elements, of which the number of non-zero elements is small:
\begin{equation}
        \begin{array}{c}      \\
                               \cdots   \\
                                    \\
                                    \end{array}
        \begin{array}{c}      \\
                               -   \\
                                    \\
                                    \end{array}
        \begin{array}{c} |     \\
                               D \\
                               |     \\
                                    \end{array}
        \begin{array}{c}      \\
                               -   \\
                                    \\
                                    \end{array}
        \begin{array}{c} |     \\
                               D   \\
                               |     \\
                                    \end{array}
        \begin{array}{c}      \\
                               -   \\
                                    \\
                                    \end{array}
       \begin{array}{c} |     \\
                               D   \\
                               |     \\
                                    \end{array}
        \begin{array}{c}      \\
                               -  \\
                                    \\
                                    \end{array}
        \begin{array}{c}      \\
                               \cdots   \\
                                    \\
                                    \end{array}
\label{eq:MPO}
\end{equation}
It is remarkable that the so-called fermionic matrix product states \cite{Bultinck}
are introduced in a similar fashion of doubling the MPO.
We have already tested this MPO approach \cite{Chung19}, and the numerical results 
by doubling the MPO were compared with those obtained by the Jordan-Wigner transformation
in good agreement.

By adopting the two-site iDMRG algorithm, we optimize two tensors in our MPS, $A_{ab}^{\sigma}$ and $B_{ab}^{\sigma}$. The physical index $\sigma$ 
takes a value of $0$ to $3$, and the indices $a$ (left) and $b$ (right) run
from $0$ to $\chi-1$, where $\chi$ is the internal bond dimension. The
Schmidt coefficients are denoted by $\lambda^{AB}_{b}$ between $A_{ab}^{\sigma}$ and $B_{bc}^{\rho}$,
and by $\lambda^{BA}_{b}$ between $B_{ab}^{\sigma}$ and $A_{bc}^{\rho}$. 
A state in the MPS space is represented by graphic notation such as
\begin{equation}
        \begin{array}{c}      \\
                               \cdots   \\
                                    \end{array}
        \begin{array}{c}      \\
                               -   \\
                                    \end{array}
        \begin{array}{c} |     \\
                               A \\
                                    \end{array}
        \begin{array}{c}      \\
                               -   \\
                                    \end{array}
        \begin{array}{c}      \\
                               \lambda^{AB}   \\
                                    \end{array}
        \begin{array}{c}      \\
                               -   \\
                                    \end{array}
        \begin{array}{c} |     \\
                               
                               B   \\
                                    \end{array}
        \begin{array}{c}      \\
                               -   \\
                                    \end{array}
        \begin{array}{c}      \\
                               \lambda^{BA}   \\
                                    \end{array}
        \begin{array}{c}      \\
                               -   \\
                                    \end{array}
       \begin{array}{c} |     \\
                               A   \\
                                    \end{array}
        \begin{array}{c}      \\
                               -   \\
                                    \end{array}
        \begin{array}{c}      \\
                               \cdots   \\
                                    \end{array}
\label{eq:MPS}
\end{equation}

By evaluating $4 \times 12 \times 12 \times 4$ elements of $D$, we perform iDMRG by acting
the MPO of Eq. (\ref{eq:MPO}) on the MPS of Eq. (\ref{eq:MPS}).
Regardless of the values of $t$, $U$, $\mu_{e}$, and $\mu_{o}$ used in our calculations, we have
observed a smooth convergence. Our numerical iDMRG results show that
the ground-state solutions are divided into two classes: the MPS are either of the form
$\cdots ABABAB \cdots$, which we will identify as a Mott
phase with unit cell of two sites, near half filling, or uniform $\cdots AAAAAA \cdots$, 
which we will identify as metallic, further away from half filling.

\section{Numerical Results}

By setting $t=1$ and $U=4$, for a given $\mu_{e}$ and $\mu_{o}$, we calculate 
the ground state by the two-site iDMRG. With the ground state, we extract three
physical quantities corresponding to even and odd sites: the bond 
energy $\langle h_{e} \rangle$ and $\langle h_{o} \rangle$,
the number density $\langle n_{e} \rangle$ and
$\langle n_{o} \rangle$, and the half-chain entanglement entropy $S_{e}$ and $S_{o}$.
Here, the bond Hamiltonian $h_{e}~(i=\text{even})$ is written as
\begin{eqnarray}
h_{e} &=& -t(c^{\dagger}_{i\uparrow}c_{j\uparrow} +
c^{\dagger}_{j\uparrow}c_{i\uparrow} +
c^{\dagger}_{i\downarrow}c_{j\downarrow} +
c^{\dagger}_{j\downarrow}c_{i\downarrow}) \nonumber \\
& &
+\frac{U}{2}(n_{i\uparrow}-\frac{1}{2})(n_{i\downarrow}-\frac{1}{2})
-\frac{\mu_{e}}{2}
(n_{i\uparrow}+n_{i\downarrow}) \nonumber \\
& &
+\frac{U}{2}(n_{j\uparrow}-\frac{1}{2})(n_{j\downarrow}-\frac{1}{2})
-\frac{\mu_{o}}{2} (n_{j\uparrow}+n_{j\downarrow}) . \label{eq:h}
\end{eqnarray}
Also, the occupation number density operator $n_{e}~(i=\text{even})$ is written as
\begin{equation}
n_{e} = n_{i\uparrow}+n_{i\downarrow} =
c^{\dagger}_{i\uparrow}c_{i\uparrow} + c^{\dagger}_{i\downarrow}c_{i\downarrow} .
\end{equation}
Using the Schmidt coefficients $\lambda_{a}$ between $A$ and $B$ in the MPS, we determine
$S_{e}$. Keeping the normalization with $\sum^{\chi - 1}_{a=0}\lambda^{2}_{a}=1$, we
calculate the half-chain entanglement entropy $S_{e}$, which is
given by
\begin{equation}
S_{e} = -\sum^{\chi - 1}_{a=0}\lambda^{2}_{a} \log_{2}
\lambda^{2}_{a}. \label{eq:EE2}
\end{equation}
Obviously, we choose $i=\text{odd}$ for $h_{o}$ and $n_{o}$. We use
the Schmidt coefficients $\lambda^{BA}$ between $B$ and $A$ in the MPS for $S_{o}$.

\begin{figure}
\includegraphics[width= 9.0 cm]{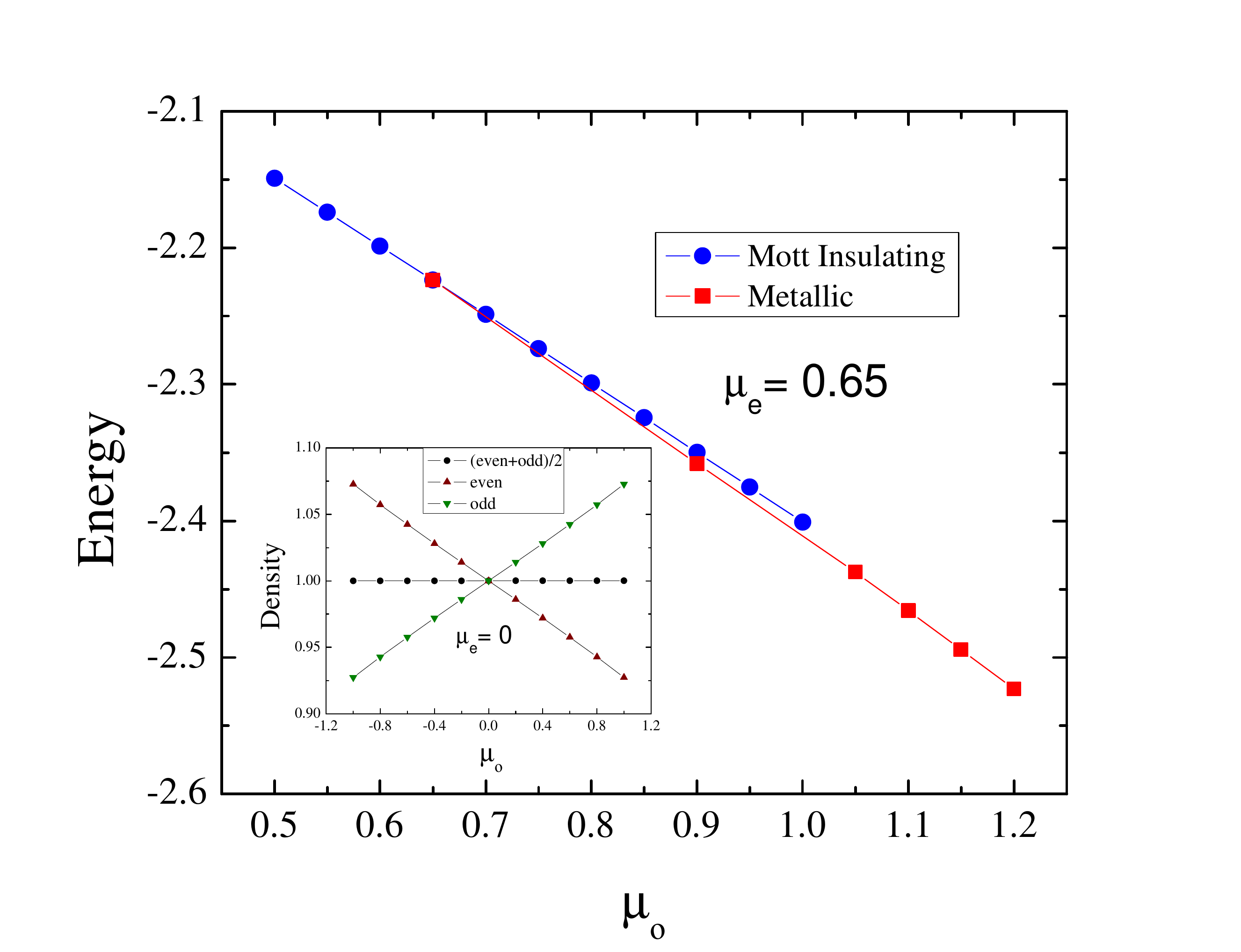}
\caption {Ground-state bond energy $(\langle h_{e} \rangle + \langle h_{o} \rangle)/2$ 
versus $\mu_{o}$ showing a level-crossing between
Mott-insulating and metallic solutions, at $\mu_{e}=0.65$ and $\chi = 100$. We set
$t = 1$ and $U = 4$. Inset: the number density of $\langle n_{e} \rangle$, $\langle n_{o} \rangle$ 
and $(\langle n_{e} \rangle + \langle n_{o} \rangle)/2$ 
versus $\mu_{o}$ showing the Mott-insulating states, at $\mu_{e}=0$ and $\chi = 100$.  
Although the difference of $(\langle n_{e} \rangle - \langle n_{o} \rangle)$ increases 
as the staggered potential $\Delta  = (\mu_{o} - \mu_{e})$ becomes bigger, 
the average number density $(\langle n_{e} \rangle + \langle n_{o} \rangle)/2$ 
remains fixed as $1$
in the Mott-insulating states.
\label{fig:fig2}  }
\end{figure}

In order to verify the correctness of the calculation, we use
an important analytic result \cite{Essler} from the critical
point of $\mu_{+}$ at which the Mott insulator - metallic phase
transition takes place in the restricted parameter space of 
$\mu_{e} = \mu_{o}$ or $\Delta = 0$. In fact, the transition point $\mu_{+}$ is given by
\begin{equation}
\mu_{+} = -2 + \frac{U}{2} + 2  \int_{0}^{\infty} \frac{d
\omega}{\omega}\frac{J_{1}(\omega)\exp(-\omega U /4)}{\cosh(\omega
U/4)}. \label{eq:mu}
\end{equation}
For $U = 4$, we obtain the exact critical point, $\mu_{+} = \mu_{e} = \mu_{o} = 0.643364$.
We present the average bond energy $(\langle h_{e} \rangle
+\langle h_{o} \rangle)/2$ as shown in Fig. \ref{fig:fig2}, where the level-crossing takes place at $\mu_{e} = \mu_{o} \sim 0.65$.

In Fig. \ref{fig:fig2}, to determine the level-crossing of the average bond energy,
we compute the ground state starting from $\mu_{o}=0$. On the other hand, we can also
calculate the ground state of the uniform solution by
decreasing $\mu_{o}$ from large values. In each iDMRG
calculation, the tensors of the initial environment are given by
the previous solution of the different $\mu_{o}$.

\begin{figure}
\includegraphics[width= 9.0 cm]{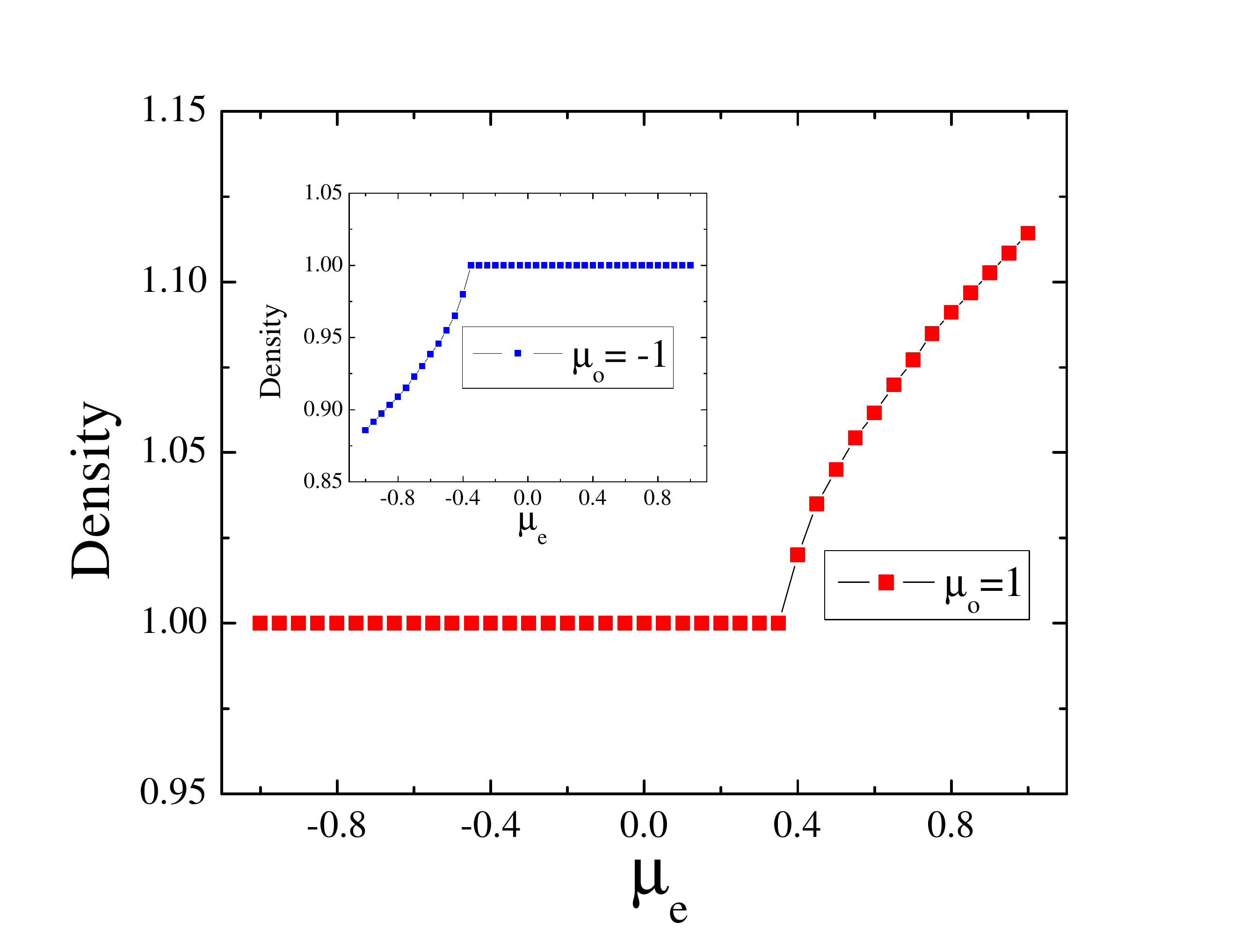}
\caption {The number density $(\langle n_{e} \rangle + \langle n_{o} \rangle)/2$ versus $\mu_{e}$ showing the abrupt change
near $\mu_{e}=0.35$, at $\mu_{o}=1$, by setting $t=1$, $U=4$ and $\chi=100$. 
We note that the phase transition takes place near $\mu_{e}=0.35$. 
Inset: By setting $\mu_{o}=-1$, the number density versus $\mu_{e}$ 
showing the phase transition near $\mu_{e}=-0.35$.
\label{fig:fig3}  }
\end{figure}

\begin{figure}
\includegraphics[width= 9.0 cm]{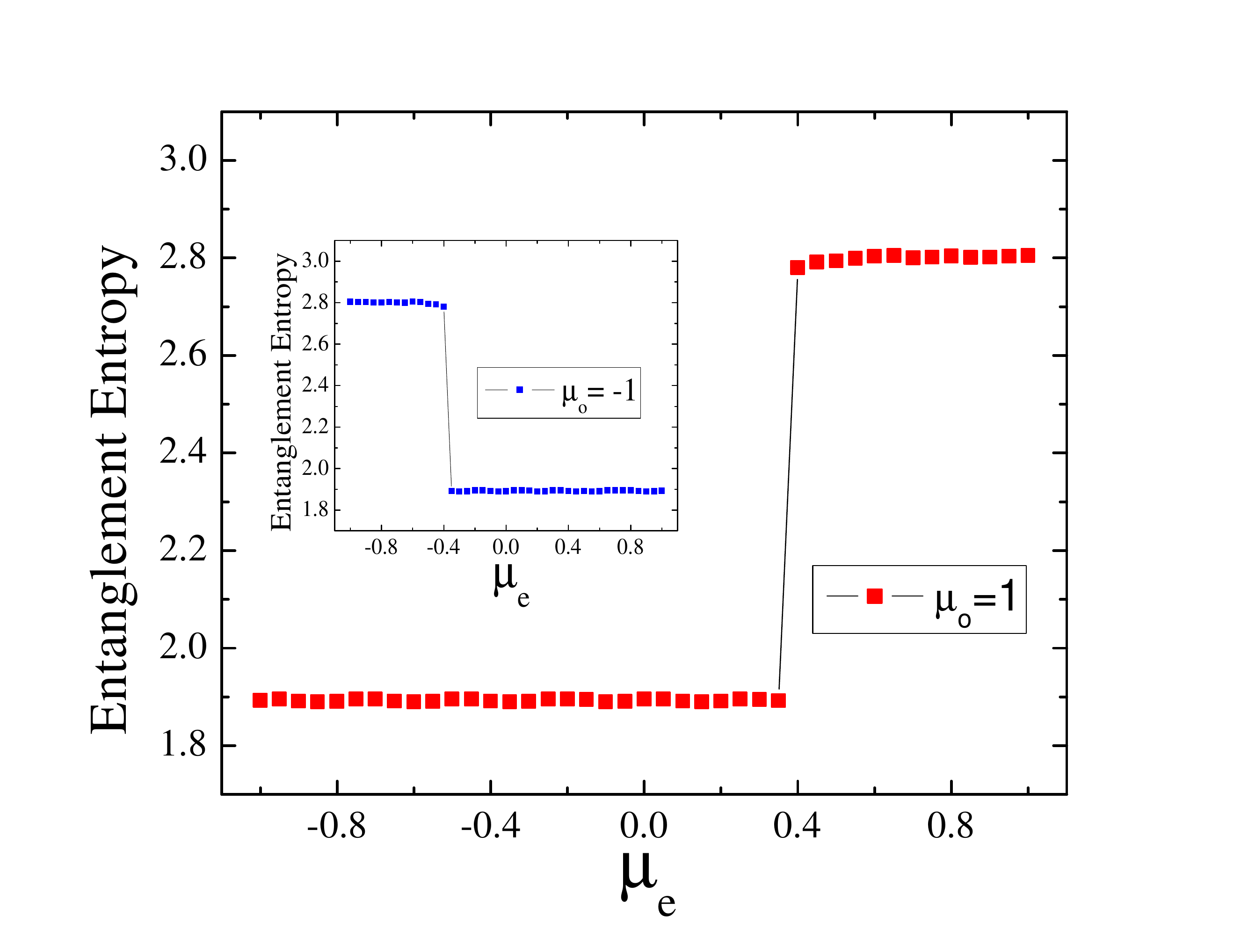}
\caption {The half-chain entanglement entropy $(S_{e}+S_{o})/2$ versus $\mu_{e}$ showing the jump 
near $\mu_{e}=0.35$, at $\mu_{o}=1$, by setting $t=1$, $U=4$ and $\chi=100$. 
We note that the phase transition takes place near $\mu_{e}=0.35$. 
Inset: By setting $\mu_{o}=-1$, the half-chain entanglement entropy versus $\mu_{e}$ showing the phase transition near $\mu_{e}=-0.35$.
\label{fig:fig4}  }
\end{figure}

In order to determine the Mott transition in the parameter space of 
$\mu_{e}$ and $\mu_{o}$, we compute the average number density 
$(\langle n_{e} \rangle + \langle n_{o} \rangle)/2$ and the average entanglement entropy $(S_{e}+S_{o})/2$.
Fixing $\mu_{o}$, we present $( \langle n_{e} \rangle + \langle n_{o} \rangle )/2$ and $(S_{e}+S_{o})/2$ by varying $\mu_{e}$,
as shown in Fig. \ref{fig:fig3} and \ref{fig:fig4}.
We note that the abrupt behaviour corresponds to the Mott transition.

\begin{figure}
\includegraphics[width= 9.0 cm]{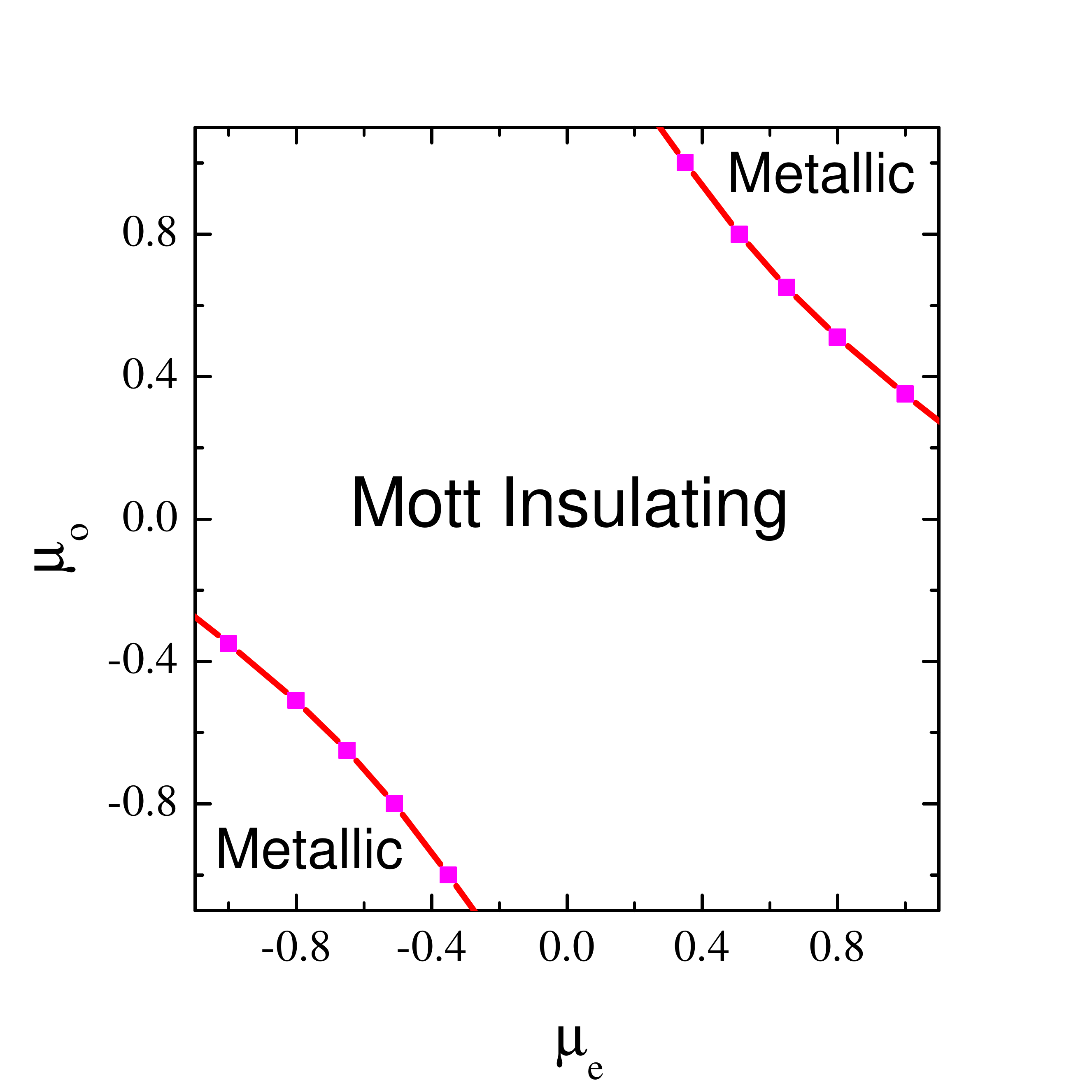}
\caption {Numerical phase diagram of the one-dimensional ionic Hubbard model. We have set $t = 1$
and $U=4$. In the region of the Mott-insulating phase, the number density is given by $1$, and the entanglement entropy is almost a constant. The red
line is a guide to the eyes. The region of the Mott-insulating phase shrinks for smaller $U$. \label{fig:fig5} }
\end{figure}

As the main numerical result, we present the phase diagram in Fig. \ref{fig:fig5}.
One notices the robustness of the Mott state. In fact, we find that
a small difference of energy offsets, which is the staggered potential
$\Delta = \mu_{o} - \mu_{e}$, can not destroy the Mott phase.

\section{Conclusion}

In summary, we have used iDMRG with MPO to obtain the ground state
in the one-dimensional ionic Hubbard model.
We calculate the occupation number density and the half-chain
entanglement entropy to determine the Mott transition.
We draw the phase diagram in the two-dimensional parameter space of the chemical potential and the
staggered potential.

It is of interest to extend our method to manage the number of fermions
in two-dimensional systems.
A typical topic of interest may be the two-dimensional fermion
Hubbard model, for which our first task is to build a local tensor product operator
like MPO. 
Although we do not encounter the
notorious sign problem here, we should overcome the sign problem
in the two-dimensional Hubbard model with the tensor product operator.
As we double MPO as explained above, we can double the tensor 
product operator in order to support the number of fermions. We anticipate
progress in the two-dimensional case.

\begin{acknowledgments}
This work was partially supported by the Basic Science Research
Program through the National Research Foundation of Korea (NRF)
funded by the Ministry of Education, Science and Technology (Grant
No. NRF-2017R1D1A1A0201845). The author would like to
thank M. C. Cha for helpful discussions.
\end{acknowledgments}

\nocite{*}
\bibliography{bib}

\providecommand{\noopsort}[1]{}\providecommand{\singleletter}[1]{#1}%
\begin{thebibliography}{26}%
\makeatletter
\providecommand \@ifxundefined [1]{%
 \@ifx{#1\undefined}
}%
\providecommand \@ifnum [1]{%
 \ifnum #1\expandafter \@firstoftwo
 \else \expandafter \@secondoftwo
 \fi
}%
\providecommand \@ifx [1]{%
 \ifx #1\expandafter \@firstoftwo
 \else \expandafter \@secondoftwo
 \fi
}%
\providecommand \natexlab [1]{#1}%
\providecommand \enquote  [1]{``#1''}%
\providecommand \bibnamefont  [1]{#1}%
\providecommand \bibfnamefont [1]{#1}%
\providecommand \citenamefont [1]{#1}%
\providecommand \href@noop [0]{\@secondoftwo}%
\providecommand \href [0]{\begingroup \@sanitize@url \@href}%
\providecommand \@href[1]{\@@startlink{#1}\@@href}%
\providecommand \@@href[1]{\endgroup#1\@@endlink}%
\providecommand \@sanitize@url [0]{\catcode `\\12\catcode `\$12\catcode
  `\&12\catcode `\#12\catcode `\^12\catcode `\_12\catcode `\%12\relax}%
\providecommand \@@startlink[1]{}%
\providecommand \@@endlink[0]{}%
\providecommand \url  [0]{\begingroup\@sanitize@url \@url }%
\providecommand \@url [1]{\endgroup\@href {#1}{\urlprefix }}%
\providecommand \urlprefix  [0]{URL }%
\providecommand \Eprint [0]{\href }%
\providecommand \doibase [0]{https://doi.org/}%
\providecommand \selectlanguage [0]{\@gobble}%
\providecommand \bibinfo  [0]{\@secondoftwo}%
\providecommand \bibfield  [0]{\@secondoftwo}%
\providecommand \translation [1]{[#1]}%
\providecommand \BibitemOpen [0]{}%
\providecommand \bibitemStop [0]{}%
\providecommand \bibitemNoStop [0]{.\EOS\space}%
\providecommand \EOS [0]{\spacefactor3000\relax}%
\providecommand \BibitemShut  [1]{\csname bibitem#1\endcsname}%
\let\auto@bib@innerbib\@empty
\bibitem [{\citenamefont {Osterloh}\ \emph {et~al.}(2002)\citenamefont
  {Osterloh}, \citenamefont {Amico}, \citenamefont {Falci},\ and\ \citenamefont
  {Fazio}}]{Osterloh}%
  \BibitemOpen
  \bibfield  {author} {\bibinfo {author} {\bibfnamefont {A.}~\bibnamefont
  {Osterloh}}, \bibinfo {author} {\bibfnamefont {L.}~\bibnamefont {Amico}},
  \bibinfo {author} {\bibfnamefont {G.}~\bibnamefont {Falci}},\ and\ \bibinfo
  {author} {\bibfnamefont {R.}~\bibnamefont {Fazio}},\ }\href@noop {}
  {\bibfield  {journal} {\bibinfo  {journal} {Nature}\ }\textbf {\bibinfo
  {volume} {416}},\ \bibinfo {pages} {608} (\bibinfo {year}
  {2002})}\BibitemShut {NoStop}%
\bibitem [{\citenamefont {Osborne}\ and\ \citenamefont
  {Nielsen}(2002)}]{Osborne}%
  \BibitemOpen
  \bibfield  {author} {\bibinfo {author} {\bibfnamefont {T.~J.}\ \bibnamefont
  {Osborne}}\ and\ \bibinfo {author} {\bibfnamefont {M.~A.}\ \bibnamefont
  {Nielsen}},\ }\href@noop {} {\bibfield  {journal} {\bibinfo  {journal}
  {Phys.\ Rev.\ A}\ }\textbf {\bibinfo {volume} {66}},\ \bibinfo {pages}
  {032110} (\bibinfo {year} {2002})}\BibitemShut {NoStop}%
\bibitem [{\citenamefont {Sachdev}(2011)}]{Sachdev}%
  \BibitemOpen
  \bibfield  {author} {\bibinfo {author} {\bibfnamefont {S.}~\bibnamefont
  {Sachdev}},\ }\href@noop {} {\emph {\bibinfo {title} {Quantum Phase
  Transitions}}}\ (\bibinfo  {publisher} {Cambridge University Press},\
  \bibinfo {year} {2011})\BibitemShut {NoStop}%
\bibitem [{\citenamefont {Eisert}\ \emph {et~al.}(2010)\citenamefont {Eisert},
  \citenamefont {Cramer},\ and\ \citenamefont {Plenio}}]{Eisert}%
  \BibitemOpen
  \bibfield  {author} {\bibinfo {author} {\bibfnamefont {J.}~\bibnamefont
  {Eisert}}, \bibinfo {author} {\bibfnamefont {M.}~\bibnamefont {Cramer}},\
  and\ \bibinfo {author} {\bibfnamefont {M.~B.}\ \bibnamefont {Plenio}},\
  }\href@noop {} {\bibfield  {journal} {\bibinfo  {journal} {Rev.\ Mod.\
  Phys.}\ }\textbf {\bibinfo {volume} {82}},\ \bibinfo {pages} {277} (\bibinfo
  {year} {2010})}\BibitemShut {NoStop}%
\bibitem [{\citenamefont {Cha}\ and\ \citenamefont {Chung}(2018)}]{Cha}%
  \BibitemOpen
  \bibfield  {author} {\bibinfo {author} {\bibfnamefont {M.~C.}\ \bibnamefont
  {Cha}}\ and\ \bibinfo {author} {\bibfnamefont {M.~H.}\ \bibnamefont
  {Chung}},\ }\href@noop {} {\bibfield  {journal} {\bibinfo  {journal} {Physica
  \ B}\ }\textbf {\bibinfo {volume} {536}},\ \bibinfo {pages} {701} (\bibinfo
  {year} {2018})}\BibitemShut {NoStop}%
\bibitem [{\citenamefont {Tagliacozzo}\ \emph {et~al.}(2008)\citenamefont
  {Tagliacozzo}, \citenamefont {de~Oliveira}, \citenamefont {Iblisdir},\ and\
  \citenamefont {Latorre}}]{Tagliacozzo}%
  \BibitemOpen
  \bibfield  {author} {\bibinfo {author} {\bibfnamefont {L.}~\bibnamefont
  {Tagliacozzo}}, \bibinfo {author} {\bibfnamefont {T.~R.}\ \bibnamefont
  {de~Oliveira}}, \bibinfo {author} {\bibfnamefont {S.}~\bibnamefont
  {Iblisdir}},\ and\ \bibinfo {author} {\bibfnamefont {J.~I.}\ \bibnamefont
  {Latorre}},\ }\href@noop {} {\bibfield  {journal} {\bibinfo  {journal}
  {Phys.\ Rev.\ B}\ }\textbf {\bibinfo {volume} {78}},\ \bibinfo {pages}
  {024410} (\bibinfo {year} {2008})}\BibitemShut {NoStop}%
\bibitem [{\citenamefont {Pollmann}\ \emph {et~al.}(2010)\citenamefont
  {Pollmann}, \citenamefont {Turner}, \citenamefont {Berg},\ and\ \citenamefont
  {Oshikawa}}]{Pollmann}%
  \BibitemOpen
  \bibfield  {author} {\bibinfo {author} {\bibfnamefont {F.}~\bibnamefont
  {Pollmann}}, \bibinfo {author} {\bibfnamefont {A.~M.}\ \bibnamefont
  {Turner}}, \bibinfo {author} {\bibfnamefont {E.}~\bibnamefont {Berg}},\ and\
  \bibinfo {author} {\bibfnamefont {M.}~\bibnamefont {Oshikawa}},\ }\href@noop
  {} {\bibfield  {journal} {\bibinfo  {journal} {Phys.\ Rev.\ B}\ }\textbf
  {\bibinfo {volume} {81}},\ \bibinfo {pages} {064439} (\bibinfo {year}
  {2010})}\BibitemShut {NoStop}%
\bibitem [{\citenamefont {Pirvu}\ \emph {et~al.}(2012)\citenamefont {Pirvu},
  \citenamefont {Vidal}, \citenamefont {Verstraete},\ and\ \citenamefont
  {Tagliacozzo}}]{Pirvu}%
  \BibitemOpen
  \bibfield  {author} {\bibinfo {author} {\bibfnamefont {B.}~\bibnamefont
  {Pirvu}}, \bibinfo {author} {\bibfnamefont {G.}~\bibnamefont {Vidal}},
  \bibinfo {author} {\bibfnamefont {F.}~\bibnamefont {Verstraete}},\ and\
  \bibinfo {author} {\bibfnamefont {L.}~\bibnamefont {Tagliacozzo}},\
  }\href@noop {} {\bibfield  {journal} {\bibinfo  {journal} {Phys.\ Rev.\ B}\
  }\textbf {\bibinfo {volume} {86}},\ \bibinfo {pages} {075117} (\bibinfo
  {year} {2012})}\BibitemShut {NoStop}%
\bibitem [{\citenamefont {Pino}\ \emph {et~al.}(2012)\citenamefont {Pino},
  \citenamefont {Prior}, \citenamefont {Somoza}, \citenamefont {Jaksch},\ and\
  \citenamefont {Clark}}]{Pino}%
  \BibitemOpen
  \bibfield  {author} {\bibinfo {author} {\bibfnamefont {M.}~\bibnamefont
  {Pino}}, \bibinfo {author} {\bibfnamefont {J.}~\bibnamefont {Prior}},
  \bibinfo {author} {\bibfnamefont {A.~M.}\ \bibnamefont {Somoza}}, \bibinfo
  {author} {\bibfnamefont {D.}~\bibnamefont {Jaksch}},\ and\ \bibinfo {author}
  {\bibfnamefont {S.~R.}\ \bibnamefont {Clark}},\ }\href@noop {} {\bibfield
  {journal} {\bibinfo  {journal} {Phys.\ Rev.\ A}\ }\textbf {\bibinfo {volume}
  {86}},\ \bibinfo {pages} {023631} (\bibinfo {year} {2012})}\BibitemShut
  {NoStop}%
\bibitem [{\citenamefont {Gu}\ \emph {et~al.}(2004)\citenamefont {Gu},
  \citenamefont {Deng}, \citenamefont {Li},\ and\ \citenamefont {Lin}}]{Gu}%
  \BibitemOpen
  \bibfield  {author} {\bibinfo {author} {\bibfnamefont {S.-J.}\ \bibnamefont
  {Gu}}, \bibinfo {author} {\bibfnamefont {S.-S.}\ \bibnamefont {Deng}},
  \bibinfo {author} {\bibfnamefont {Y.-Q.}\ \bibnamefont {Li}},\ and\ \bibinfo
  {author} {\bibfnamefont {H.-Q.}\ \bibnamefont {Lin}},\ }\href@noop {}
  {\bibfield  {journal} {\bibinfo  {journal} {Phys.\ Rev.\ Lett.}\ }\textbf
  {\bibinfo {volume} {93}},\ \bibinfo {pages} {086402} (\bibinfo {year}
  {2004})}\BibitemShut {NoStop}%
\bibitem [{\citenamefont {Larsson}\ and\ \citenamefont
  {Johannesson}(2005)}]{Larsson}%
  \BibitemOpen
  \bibfield  {author} {\bibinfo {author} {\bibfnamefont {D.}~\bibnamefont
  {Larsson}}\ and\ \bibinfo {author} {\bibfnamefont {H.}~\bibnamefont
  {Johannesson}},\ }\href@noop {} {\bibfield  {journal} {\bibinfo  {journal}
  {Phys.\ Rev.\ Lett.}\ }\textbf {\bibinfo {volume} {95}},\ \bibinfo {pages}
  {196406} (\bibinfo {year} {2005})}\BibitemShut {NoStop}%
\bibitem [{\citenamefont {Iemini}\ \emph {et~al.}(2015)\citenamefont {Iemini},
  \citenamefont {Maciel},\ and\ \citenamefont {Vianna}}]{Iemini}%
  \BibitemOpen
  \bibfield  {author} {\bibinfo {author} {\bibfnamefont {F.}~\bibnamefont
  {Iemini}}, \bibinfo {author} {\bibfnamefont {T.~O.}\ \bibnamefont {Maciel}},\
  and\ \bibinfo {author} {\bibfnamefont {R.~O.}\ \bibnamefont {Vianna}},\
  }\href@noop {} {\bibfield  {journal} {\bibinfo  {journal} {Phys.\ Rev.\ B}\
  }\textbf {\bibinfo {volume} {92}},\ \bibinfo {pages} {075423} (\bibinfo
  {year} {2015})}\BibitemShut {NoStop}%
\bibitem [{\citenamefont {Cha}(2018)}]{Cha18}%
  \BibitemOpen
  \bibfield  {author} {\bibinfo {author} {\bibfnamefont {M.~C.}\ \bibnamefont
  {Cha}},\ }\href@noop {} {\bibfield  {journal} {\bibinfo  {journal} {Phys.\
  Rev.\ B}\ }\textbf {\bibinfo {volume} {98}},\ \bibinfo {pages} {235161}
  (\bibinfo {year} {2018})}\BibitemShut {NoStop}%
\bibitem [{\citenamefont {Parsons}\ \emph {et~al.}(2015)\citenamefont
  {Parsons}, \citenamefont {Huber}, \citenamefont {Mazurenko}, \citenamefont
  {Chiu}, \citenamefont {Setiawan}, \citenamefont {Wooley-Brown},\ and\
  \citenamefont {S.~Blatt}}]{Parsons}%
  \BibitemOpen
  \bibfield  {author} {\bibinfo {author} {\bibfnamefont {M.~F.}\ \bibnamefont
  {Parsons}}, \bibinfo {author} {\bibfnamefont {F.}~\bibnamefont {Huber}},
  \bibinfo {author} {\bibfnamefont {A.}~\bibnamefont {Mazurenko}}, \bibinfo
  {author} {\bibfnamefont {C.~S.}\ \bibnamefont {Chiu}}, \bibinfo {author}
  {\bibfnamefont {W.}~\bibnamefont {Setiawan}}, \bibinfo {author}
  {\bibfnamefont {K.}~\bibnamefont {Wooley-Brown}},\ and\ \bibinfo {author}
  {\bibfnamefont {M.~G.}\ \bibnamefont {S.~Blatt}},\ }\href@noop {} {\bibfield
  {journal} {\bibinfo  {journal} {Phys.\ Rev.\ Lett.}\ }\textbf {\bibinfo
  {volume} {114}},\ \bibinfo {pages} {213002} (\bibinfo {year}
  {2015})}\BibitemShut {NoStop}%
\bibitem [{\citenamefont {Cheuk}\ \emph {et~al.}(2015)\citenamefont {Cheuk},
  \citenamefont {Nichols}, \citenamefont {Okan}, \citenamefont {Gersdorf},
  \citenamefont {Ramasesh}, \citenamefont {Bakr}, \citenamefont {Lompe},\ and\
  \citenamefont {Zwierlein}}]{Cheuk}%
  \BibitemOpen
  \bibfield  {author} {\bibinfo {author} {\bibfnamefont {L.~W.}\ \bibnamefont
  {Cheuk}}, \bibinfo {author} {\bibfnamefont {M.~A.}\ \bibnamefont {Nichols}},
  \bibinfo {author} {\bibfnamefont {M.}~\bibnamefont {Okan}}, \bibinfo {author}
  {\bibfnamefont {T.}~\bibnamefont {Gersdorf}}, \bibinfo {author}
  {\bibfnamefont {V.~V.}\ \bibnamefont {Ramasesh}}, \bibinfo {author}
  {\bibfnamefont {W.~S.}\ \bibnamefont {Bakr}}, \bibinfo {author}
  {\bibfnamefont {T.}~\bibnamefont {Lompe}},\ and\ \bibinfo {author}
  {\bibfnamefont {M.~W.}\ \bibnamefont {Zwierlein}},\ }\href@noop {} {\bibfield
   {journal} {\bibinfo  {journal} {Phys.\ Rev.\ Lett.}\ }\textbf {\bibinfo
  {volume} {114}},\ \bibinfo {pages} {193001} (\bibinfo {year}
  {2015})}\BibitemShut {NoStop}%
\bibitem [{\citenamefont {Schreiber}\ \emph {et~al.}(2015)\citenamefont
  {Schreiber}, \citenamefont {Hodgman}, \citenamefont {Bordia}, \citenamefont
  {L{\"{u}}schen}, \citenamefont {Fischer}, \citenamefont {Vosk}, \citenamefont
  {Altman}, \citenamefont {Schneider},\ and\ \citenamefont
  {Bloch}}]{Schreiber}%
  \BibitemOpen
  \bibfield  {author} {\bibinfo {author} {\bibfnamefont {M.}~\bibnamefont
  {Schreiber}}, \bibinfo {author} {\bibfnamefont {S.~S.}\ \bibnamefont
  {Hodgman}}, \bibinfo {author} {\bibfnamefont {P.}~\bibnamefont {Bordia}},
  \bibinfo {author} {\bibfnamefont {H.~P.}\ \bibnamefont {L{\"{u}}schen}},
  \bibinfo {author} {\bibfnamefont {M.~H.}\ \bibnamefont {Fischer}}, \bibinfo
  {author} {\bibfnamefont {R.}~\bibnamefont {Vosk}}, \bibinfo {author}
  {\bibfnamefont {E.}~\bibnamefont {Altman}}, \bibinfo {author} {\bibfnamefont
  {U.}~\bibnamefont {Schneider}},\ and\ \bibinfo {author} {\bibfnamefont
  {I.}~\bibnamefont {Bloch}},\ }\href@noop {} {\bibfield  {journal} {\bibinfo
  {journal} {Science}\ }\textbf {\bibinfo {volume} {21}},\ \bibinfo {pages}
  {842} (\bibinfo {year} {2015})}\BibitemShut {NoStop}%
\bibitem [{\citenamefont {Essler}\ \emph {et~al.}(2005)\citenamefont {Essler},
  \citenamefont {Frahm}, \citenamefont {G{\"{o}}hmann}, \citenamefont
  {Kl{\"{u}}mper},\ and\ \citenamefont {Korepin}}]{Essler}%
  \BibitemOpen
  \bibfield  {author} {\bibinfo {author} {\bibfnamefont {F.~H.~L.}\
  \bibnamefont {Essler}}, \bibinfo {author} {\bibfnamefont {H.}~\bibnamefont
  {Frahm}}, \bibinfo {author} {\bibfnamefont {F.}~\bibnamefont
  {G{\"{o}}hmann}}, \bibinfo {author} {\bibfnamefont {A.}~\bibnamefont
  {Kl{\"{u}}mper}},\ and\ \bibinfo {author} {\bibfnamefont {V.~E.}\
  \bibnamefont {Korepin}},\ }\href@noop {} {\emph {\bibinfo {title} {The
  one-dimensional Hubbard model}}}\ (\bibinfo  {publisher} {Cambridge
  University Press},\ \bibinfo {year} {2005})\BibitemShut {NoStop}%
\bibitem [{\citenamefont {Torbati}\ \emph {et~al.}(2014)\citenamefont
  {Torbati}, \citenamefont {Drescher},\ and\ \citenamefont {Uhrig}}]{Torbati}%
  \BibitemOpen
  \bibfield  {author} {\bibinfo {author} {\bibfnamefont {M.~H.}\ \bibnamefont
  {Torbati}}, \bibinfo {author} {\bibfnamefont {N.~A.}\ \bibnamefont
  {Drescher}},\ and\ \bibinfo {author} {\bibfnamefont {G.~S.}\ \bibnamefont
  {Uhrig}},\ }\href@noop {} {\bibfield  {journal} {\bibinfo  {journal} {Phys.\
  Rev.\ B}\ }\textbf {\bibinfo {volume} {89}},\ \bibinfo {pages} {245126}
  (\bibinfo {year} {2014})}\BibitemShut {NoStop}%
\bibitem [{\citenamefont {Bag}\ \emph {et~al.}(2015)\citenamefont {Bag},
  \citenamefont {Garg},\ and\ \citenamefont {Krishnamurthy}}]{Bag}%
  \BibitemOpen
  \bibfield  {author} {\bibinfo {author} {\bibfnamefont {S.}~\bibnamefont
  {Bag}}, \bibinfo {author} {\bibfnamefont {A.}~\bibnamefont {Garg}},\ and\
  \bibinfo {author} {\bibfnamefont {H.~R.}\ \bibnamefont {Krishnamurthy}},\
  }\href@noop {} {\bibfield  {journal} {\bibinfo  {journal} {Phys.\ Rev.\ B}\
  }\textbf {\bibinfo {volume} {91}},\ \bibinfo {pages} {235108} (\bibinfo
  {year} {2015})}\BibitemShut {NoStop}%
\bibitem [{\citenamefont {Lin}\ \emph {et~al.}(2015)\citenamefont {Lin},
  \citenamefont {Liu}, \citenamefont {Tao},\ and\ \citenamefont {Liu}}]{Lin}%
  \BibitemOpen
  \bibfield  {author} {\bibinfo {author} {\bibfnamefont {H.~F.}\ \bibnamefont
  {Lin}}, \bibinfo {author} {\bibfnamefont {H.~D.}\ \bibnamefont {Liu}},
  \bibinfo {author} {\bibfnamefont {H.~S.}\ \bibnamefont {Tao}},\ and\ \bibinfo
  {author} {\bibfnamefont {W.~M.}\ \bibnamefont {Liu}},\ }\href@noop {}
  {\bibfield  {journal} {\bibinfo  {journal} {Scientific Reports}\ }\textbf
  {\bibinfo {volume} {5}},\ \bibinfo {pages} {9810} (\bibinfo {year}
  {2015})}\BibitemShut {NoStop}%
\bibitem [{\citenamefont {Bouadim}\ \emph {et~al.}(2007)\citenamefont
  {Bouadim}, \citenamefont {Paris}, \citenamefont {Hebert}, \citenamefont
  {Batrouni},\ and\ \citenamefont {Scalettar}}]{Bouadim}%
  \BibitemOpen
  \bibfield  {author} {\bibinfo {author} {\bibfnamefont {K.}~\bibnamefont
  {Bouadim}}, \bibinfo {author} {\bibfnamefont {N.}~\bibnamefont {Paris}},
  \bibinfo {author} {\bibfnamefont {F.}~\bibnamefont {Hebert}}, \bibinfo
  {author} {\bibfnamefont {G.~G.}\ \bibnamefont {Batrouni}},\ and\ \bibinfo
  {author} {\bibfnamefont {R.~T.}\ \bibnamefont {Scalettar}},\ }\href@noop {}
  {\bibfield  {journal} {\bibinfo  {journal} {Phys.\ Rev.\ B}\ }\textbf
  {\bibinfo {volume} {76}},\ \bibinfo {pages} {085112} (\bibinfo {year}
  {2007})}\BibitemShut {NoStop}%
\bibitem [{\citenamefont {Kampf}\ \emph {et~al.}(2003)\citenamefont {Kampf},
  \citenamefont {Sekania}, \citenamefont {Japaridze},\ and\ \citenamefont
  {Brune}}]{Kampf}%
  \BibitemOpen
  \bibfield  {author} {\bibinfo {author} {\bibfnamefont {A.~P.}\ \bibnamefont
  {Kampf}}, \bibinfo {author} {\bibfnamefont {M.}~\bibnamefont {Sekania}},
  \bibinfo {author} {\bibfnamefont {G.~I.}\ \bibnamefont {Japaridze}},\ and\
  \bibinfo {author} {\bibfnamefont {P.}~\bibnamefont {Brune}},\ }\href@noop {}
  {\bibfield  {journal} {\bibinfo  {journal} {J of Phys.: Condensed Matter}\
  }\textbf {\bibinfo {volume} {15}},\ \bibinfo {pages} {5895} (\bibinfo {year}
  {2003})}\BibitemShut {NoStop}%
\bibitem [{\citenamefont {McCulloch}(2008)}]{McCulloch}%
  \BibitemOpen
  \bibfield  {author} {\bibinfo {author} {\bibfnamefont {I.~P.}\ \bibnamefont
  {McCulloch}},\ }\href@noop {} {}\bibinfo {howpublished} {e-print
  arXiv:0804.2509} (\bibinfo {year} {2008})\BibitemShut {NoStop}%
\bibitem [{\citenamefont {Yang}\ \emph {et~al.}(2000)\citenamefont {Yang},
  \citenamefont {Kocharian},\ and\ \citenamefont {Chiang}}]{Yang}%
  \BibitemOpen
  \bibfield  {author} {\bibinfo {author} {\bibfnamefont {C.}~\bibnamefont
  {Yang}}, \bibinfo {author} {\bibfnamefont {A.~N.}\ \bibnamefont
  {Kocharian}},\ and\ \bibinfo {author} {\bibfnamefont {Y.~L.}\ \bibnamefont
  {Chiang}},\ }\href@noop {} {\bibfield  {journal} {\bibinfo  {journal} {J.
  Phys.: Condens. Matter}\ }\textbf {\bibinfo {volume} {12}},\ \bibinfo {pages}
  {7433} (\bibinfo {year} {2000})}\BibitemShut {NoStop}%
\bibitem [{\citenamefont {Bultinck}\ \emph {et~al.}(2017)\citenamefont
  {Bultinck}, \citenamefont {Williamson}, \citenamefont {Haegeman},\ and\
  \citenamefont {Verstraete}}]{Bultinck}%
  \BibitemOpen
  \bibfield  {author} {\bibinfo {author} {\bibfnamefont {N.}~\bibnamefont
  {Bultinck}}, \bibinfo {author} {\bibfnamefont {D.~J.}\ \bibnamefont
  {Williamson}}, \bibinfo {author} {\bibfnamefont {J.}~\bibnamefont
  {Haegeman}},\ and\ \bibinfo {author} {\bibfnamefont {F.}~\bibnamefont
  {Verstraete}},\ }\href@noop {} {\bibfield  {journal} {\bibinfo  {journal}
  {Phys.\ Rev.\ B}\ }\textbf {\bibinfo {volume} {95}},\ \bibinfo {pages}
  {075108} (\bibinfo {year} {2017})}\BibitemShut {NoStop}%
\bibitem [{\citenamefont {Chung}\ \emph {et~al.}(2019)\citenamefont {Chung},
  \citenamefont {Orignac}, \citenamefont {Poilblanc},\ and\ \citenamefont
  {Capponi}}]{Chung19}%
  \BibitemOpen
  \bibfield  {author} {\bibinfo {author} {\bibfnamefont {M.~H.}\ \bibnamefont
  {Chung}}, \bibinfo {author} {\bibfnamefont {E.}~\bibnamefont {Orignac}},
  \bibinfo {author} {\bibfnamefont {D.}~\bibnamefont {Poilblanc}},\ and\
  \bibinfo {author} {\bibfnamefont {S.}~\bibnamefont {Capponi}},\ }\href@noop
  {} {}\bibinfo {howpublished} {e-print arXiv:1912.10203} (\bibinfo {year}
  {2019})\BibitemShut {NoStop}%
\end{thebibliography}%
\bibliographystyle{apsrev4-2}

\end{document}